# Unitary-Only Quantum Theory Cannot Consistently Describe the Use of Itself: On the Frauchiger-Renner Paradox

R. E. Kastner
University of Maryland
rkastner@umd.edu

January 6, 2019

ABSTRACT. The Frauchiger-Renner Paradox is an extension of paradoxes based on the "Problem of Measurement," such as Schrödinger's Cat and Wigner's Friend. All of these paradoxes stem from assuming that quantum theory has only unitary (linear) physical dynamics, and the attendant ambiguity about what counts as a 'measurement'—i.e., the inability to account for the observation of determinate measurement outcomes from within the theory itself. This paper discusses a basic inconsistency arising in the FR scenario at a much earlier point than the derived contradiction: namely, the inconsistency inherent in treating an improper mixture (reduced density operator) as a proper, epistemic mixture. This is an illegitimate procedure that is nevertheless endemic if quantum theory is assumed to be always unitary. In contrast, under a non-unitary account of quantum state reduction yielding determinate outcomes, the use of a proper mixture for measurement results becomes legitimate, and this entire class of paradoxes cannot be mounted. The conclusion is that the real lesson of the FR paradox is that it is the unitary-only assumption that needs to be critically reassessed.

1. Introduction and Background

The Frauschiger-Renner Paradox (2018) is a much-discussed thought experiment in the recent literature on the foundations of quantum mechanics. This paper argues that the appropriate way to understand the FR paradox is as a *reductio ad absurdum* of the standard unitary-only form of quantum theory (UO QM); that is, as a logical refutation of UO QM. It also argues that various efforts to evade the *reductio* nature of the argument while retaining the unitary-only assumption lead to significant interpretive disadvantages. Thus, the thesis of this paper is that the FR paradox requires us to critically re-examine the usual assumption that quantum theory 'really' has only unitary evolution, and to take seriously the possibility of real non-unitarity in connection with measurement in quantum theory.

Before proceeding, we must clarify what we mean here by 'unitary-only quantum theory,' or UO QM. In the present context, UO QM includes 'textbook quantum theory,' that is, quantum theory that includes a 'projection postulate' (PP). The reason for discounting the PP as a legitimate theoretical form of non-unitarity is as follows. The usual unitary Schrödinger evolution of the quantum state (called "Process 2" by von Neumann) fails to lead to the empirically observed determinate single outcomes and resulting Boolean event structure. The addition of a 'projection postulate' is an *ad hoc* move that fails to bridge the explanatory gap between the linear evolution and what is observed. Indeed, the PP becomes part of the measurement problem, in that there is no principled physical account as to at what point in an interaction it should be applied. This ambiguous situation is even commonly thought to obligatory, with an attendant 'shifty split' or moveable 'Heisenberg Cut' that is assumed to be a necessary part of quantum theory. Yet it is the ambiguity of the use of the *ad hoc* PP in the face of supposedly unitary-only dynamics that leads to the class of "Schrödinger's Cat" paradoxes culminating in the FR inconsistency. In contrast, a genuinely non-unitary form of quantum theory has a quantitative physical account of the transition from the unitary to non-unitary dynamics, such that the 'Cut' is non-arbitrary. In such a theory, projection corresponding to the non-unitarity arises from quantitative physical content, as opposed to being a postulate without any accompanying physics. To be viable, a non-unitary quantum theory must unambiguously give rise to quantum state reduction or 'collapse' well before the advent of what would otherwise be the flagrant macroscopic superpositions invoked in this class of paradoxes.

Thus, for purposes of this discussion, all forms of quantum theory lacking a specific quantitative physical account of non-unitarity, including the circumstances of its onset, are considered as Unitary Only (UO). The usual textbook quantum theory, with only a 'projection postulate' and resulting 'shifty split,' is properly designated a kind of UO account, since the only quantitative physical account it provides for the quantum state is that of the unitary ("Process 2") evolution. The present author is aware that designating textbook quantum theory as UO is a different usage from some authors, who take inclusion of the PP as a non-unitary form of the theory. However, as noted above, such approaches have no physics to support the postulated non-unitarity, and as such, they fail to constitute non-unitary physical theories. The only dynamics for which they provide a specific quantitative account is the unitary dynamics. Thus, they are unitary-only, even if they help themselves to a projection postulate for ostensible (after-the-fact) consistency with the observed Boolean event structure. To distinguish 'textbook' presentations of quantum theory that include the PP (but lack any actual physics of non-unitarity) from Everettian-type theories that deny collapse, we will denote the former as UOPP.[1] It should be noted that in recent years, in view of the UO decoherence program which has often mistakenly been taken to 'explain' classical determinacy,[2] PP tends to be less fashionable, and many mainstream discussions of quantum theory eschew the idea of PP or 'reduction' in favor of an Everettian view. The ambiguity surrounding 'whether to postulate projection or not' under an assumed linear-only dynamics is also a key aspect of the FR paradox.

We now turn to specifics of the FR paradox. The paradox consists of a contradiction in which two sets of observers each apply quantum theory seemingly 'correctly,' but must ultimately disagree strongly on the probability of the outcome of a measurement on the same total system. The argument is based on an elaboration of the 'Wigner's Friend' paradox with suitably chosen states.

---

[1] Perhaps a suitable term for the formulation of quantum theory having only the PP as a token of non-unitarity is 'fig leaf non-unitarity.'

[2] The UO decoherence program has been criticized as circular; see, e.g., Kastner (2014) and references therein. In any case, the results of the standard decoherence program can be derived in the Transactional Interpretation (TI), which has real non-unitary reduction. Under TI, the decoherence functions arise from proper mixtures that can be given an epistemic interpretation, and the account of classical emergence is non-circular (Kastner 2019b).

FR(2018) lay out three assumption upon which they base their argument. They denote these Consistency (C), Single Outcome (S), and Quantum Theory's Universal Correctness (Q). They say:

> "Assumption (Q) captures the universal correctness of quantum theory (specifically, it proclaims that an agent can be certain that a given proposition holds whenever the quantum mechanical Born rule assigns probability 1 to it), (C) demands consistency in the sense illustrated by Fig. 1, [this shows observers at different levels of complexity agreeing as to the outcome of a measurement at a particular level], and (S) ensures that, from the viewpoint of an agent who carries out a measurement, the measurement has one single outcome (e.g., if the agent observes z = +½ then she can be certain that she did not also observe z = – ½ )."

Assumption Q is stated only in terms of the Born Rule. However, the key assumption in the way 'Q' is applied is that nothing actually (physically) happens during a measurement beyond the establishment of unitary correlations. This is the Unitary Only assumption (UO). The unitary evolution corresponds to von Neumann's unitary 'Process 2,' which (as noted above) is generally assumed to constitute the only *real* physical dynamics; von Neumann's 'Process 1,' involving 'collapse' or reduction, is tacitly assumed not to be a real physical process. We say 'tacitly' here, since this is not made explicit in the FR paper. However, they treat 'collapse' theories as different theories from quantum theory, since they claim (in their Table III) that such theories violate assumption Q. Thus, they assume that theories with physical non-unitarity must deviate from the Born Rule (which is not the case; see Section 3). This evidences the general belief that 'real' quantum theory must lack physical reduction of the state vector—i.e., that it must be physically UO.

I hasten to add that the unitarity assumption underlying the FR paradox is pointed out (at least) by Sudbery (2019), Nurgalieva and del Rio (2019), and Araújo (2018); I make no claim that the crucial role of unitarity in deriving the contradiction has not been noticed before. Indeed, it is only 'by linearity' that Frauchiger and Renner arrive at their 'absurd superposition' (to be discussed in more detail below). However, the thesis of the present paper is that the core of the problem 'brought to a head' by the FR paradox is precisely that the prevailing view of quantum theory (including hidden variables or 'modal' approaches) treats UO as an obligatory aspect of quantum theory, when arguably what is required for full consistency of quantum theory is that UO be relinquished. For an appropriate non-unitary theory, quantum theory can still perfectly well obey the Born Rule, thus satisfying Q as stated above. This issue is elaborated in Section 3.

FR introduce their argument by envisioning a scenario with an experimenter F, his spin-½ system S, and his measuring device D, and another experimenter W outside the lab. F measures the spin of S, finding a result. Yet it is assumed ("by linearity") that the unitary evolution still continues such that W must describe F's entire lab by an entangled state. These two levels of description are presented as follows: (we'll use letters here to distinguish them from later equation numbers):

- F assigns to his system S, after measuring its spin along z, either

$$|\uparrow\rangle_S \text{ or } |\downarrow\rangle_S \qquad (A)$$

- The external observer W's assigns to S, D, F, all of which are taken to comprise a 'lab' L, the "absurd" (their word) superposition:

$$|\Psi\rangle_L = \frac{1}{\sqrt{2}}\left(|\uparrow\rangle_S \otimes \left|"z=+\tfrac{1}{2}"\right\rangle_D \otimes |"\psi_S = |\uparrow\rangle"\rangle_F + |\downarrow\rangle_S \otimes \left|"z=-\tfrac{1}{2}"\right\rangle_D \otimes |"\psi_S = |\downarrow\rangle"\rangle_F\right) \qquad (B)$$

From these assumptions, by way of some carefully chosen initial states and interactions, FR obtain a contradiction in which observers such as F at a lower level of complexity predict zero probability for an outcome that has a non-vanishing probability to be found by postulated 'super-observers' such as W at a higher level of complexity. Bub (2017) gives a clear and accessible presentation of the details of the quantum states involved, which we will not repeat here (although this author arrives at different conclusions regarding the implications of the result). It turns out that this absurd result—the *reductio ad absurdum* of the UO account of quantum theory--is not surprising if we look at an earlier stage in the construction of the FR paradox, which contains a fundamental inconsistency in the way in which the UO account is applied.

2. The Origin of the Inconsistency

Let us review the usual account of 'measurement' under the unitary-only (UO) assumption. For an observable $R$ with eigenvalues $\{r\}$, where the system S is initially in a superposition $\psi$ of eigenstates $|r\rangle$, measurement (so the story goes) consists of nothing more than the introduction of a correlation between S and a device D, originally in a ready state $\phi_0$:

$$|\psi\rangle|\phi_0\rangle \to \sum_r a_r |r\rangle|\phi_r\rangle \qquad (1)$$

We must first note that unitary-only quantum theory does not correctly license the attribution of an eigenstate $|r\rangle$ to any system found in a definite outcome of eigenvalue $r$, since according to UO such a system continues on indefinitely, by (1), as a component subsystem of a composite entangled state. This applies to UOPP as well, since no physical account is given of 'collapse' to an outcome eigenstate. Thus, the application of PP is not physically justified, and is an *ad hoc* move. If any 'measured' system actually remains a component of an entangled state as in (1) (and is therefore actually a subsystem of a larger composite system in a pure state), the inappropriateness of applying an outcome eigenstate is no more than the elementary observation that such a subsystem is in an *improper* mixed state, as opposed to a pure state or proper, 'epistemic' mixed state that can be interpreted as representing ignorance about an actual outcome. As R. I. G. Hughes put it regarding the customary (but inadequate) account of 'measurement' under the unitary-only assumption:

> "Alas, elegant as this treatment is, as an account of the possible to the actual it just won't do. … What we would like to say, when we speak of the measurement device being in a mixture of [projectors corresponding to outcomes] is that it actually *is* in one of these pure states but we don't know which; in other words, we would like to use the ignorance interpretation of mixtures. But, as we saw in section 5.8, this interpretation cannot be used for those mixtures which arise from a reduction of a pure state in a tensor-product space [like that arising from a unitary interaction of a system with other degrees of freedom]." (Hughes 1989, p. 283)[3]

In the present context, Hughes' locution 'possible to the actual' is synonymous with 'non-Boolean to Boolean,' where a Boolean event structure obeys the usual classical probability laws that can assign determinate outcomes in a consistent manner—which the pure state (1) does not allow. Of course, the fact that empirically we always find a single outcome as a result of measurement (and thus a Boolean event structure) is the essence of the measurement problem facing unitary-only quantum theory. Under the UO assumption, the empirically observed Boolean event structure is a crucial (and problematic) *explanandum*; therefore, under UO, the observed Boolean events structure cannot be part of any *explanans*. All it can be is a 'given' that is inherently inconsistent with the theory (and for which the PP acts as a 'band-aid').

Returning now to the problem with the improper mixture: the situation can be illustrated through a typical EPR-Bell state, such as the triplet s=1, m=0 state for two electrons:

$$|\Psi\rangle = \frac{1}{\sqrt{2}}\left(|\uparrow\rangle|\downarrow\rangle + |\downarrow\rangle|\uparrow\rangle\right) \qquad (2)$$

The state (2) assigns a probability of unity that the total spin of the system will be 1.

If Alice measures electron A along the spin direction $z$, and Bob measures electron B along some arbitrary spin direction $\theta$, each could find the outcome 'up' or 'down.' If Alice and Bob were to assign the corresponding eigenstates to their measured degrees of freedom, the possible resulting state assignments would be:

$$\begin{aligned} |\uparrow\rangle_z \otimes |\uparrow\rangle_\theta \\ |\uparrow\rangle_z \otimes |\downarrow\rangle_\theta \\ |\downarrow\rangle_z \otimes |\uparrow\rangle_\theta \\ |\downarrow\rangle_z \otimes |\downarrow\rangle_\theta \end{aligned} \qquad (3)$$

where each pairing would be found with the appropriate probability (a function of the angle $\theta$). This would mean that, from the vantage point of someone, say 'Walter,' who did not know what those outcomes were, the composite system would have to be in the proper mixed state:

$$\rho = \sum_{i,j} \Pr(i,j) |i\rangle_z \langle i|_z \otimes |j\rangle_\theta \langle j|_\theta \qquad (4)$$

---

[3] Hughes notes that this point was first made by Feyerabend (1962).

where i,j ∈ {↑↓}. *But this contradicts the state (2).*

Nowadays, it is usually supposed that this inconsistency is not problematic, based on the following sort of reasoning: further entanglements of the electrons with environmental orthogonal states would eliminate significant interference effects for observables corresponding to the electron-only entanglement (2), such as total momentum. This suppression of interference is taken as licensing the idea that Walter would see, in a 'FAPP' sense,[4] what amounts to one of the collapsed possibilities (3). However, loss of interference does not equate to determinacy of outcomes, so FAPP utility doesn't remove the logical and interpretive inconsistency highlighted by Hughes. In any case, this usual loophole of appealing to entanglement of the degree of freedom of interest with 'external' degrees of freedom to suppress interference is not available for the 'super-observable' allegedly measurable by W in the FR paradox, which involves *all* the entangled degrees of freedom. Thus, according to UO, *consistently applied*, the states (3) are disallowed. Out of the starting gate, this blocks construction of the paradox, since construction of the paradox *depends on allowing some observers to assign states that they are not permitted, by the unitary-only assumption, to assign*.

Tausk (2019) also makes this point concerning unitary-only quantum theory: "What is wrong with this reasoning is that there is no justification for collapsing the quantum state after F's experiment, as W is going to perform a measurement of an operator having large interference terms with respect to the given macroscopic superposition…." Here, Tausk suggests that it is permitted to 'collapse the quantum state' after W performs his measurement, provided nobody else is going to perform a measurement of a certain kind. But again, 'performing a measurement,' in the sense of 'getting an outcome,' is undefined in a theoretical sense under UO; this is just the measurement problem.[5] Under these conditions, 'collapsing the quantum state' is just a calculational device, which is all that it can ever be under UO. *Under the unitary-only assumption, treating the quantum state as collapsed is a fiction*. It cannot accurately describe any physical system(s) but is applied only for the convenience or utility of an observer. This makes the account observer-dependent and instrumentalist (since under UO the assigned 'collapsed' states cannot refer to the system itself), and many of the extant attempts to save UO quantum theory in the face of the FR *reductio* resort to various forms of instrumentalism. Another alternative is a special sort of hidden variable approach (e.g., Sudbery 2016), but in order to evade the contradiction, that must also prohibit experimenters from assigning outcome-labeled eigenstates to systems for which they have observed outcomes, *if* a later measurement of a certain kind of observable is going to be performed (which, according to UO, is always possible).

Despite the fact that UO precludes assigning outcome-labeled eigenstates to 'measured' degrees of freedom, in the FR scenario, observers F inside the lab assign such eigenstates to their measured systems. This initial inconsistency naturally leads to the final inconsistent result. In fact, under UO, F is *no*t using quantum theory correctly. F's outcome-based state assignments contradict the composite entangled state, just as (4) contradicts (2). Concerning the inconsistent state assignments, FR say:

> "Although the state assignment [B] may appear to be "absurd," it does not logically contradict [A]. Indeed, the marginal on S is just a fully mixed state. While this is different from [B], the difference can be explained by the agents' distinct level of knowledge: F has observed z and hence knows the spin direction, whereas W is ignorant about it."

---

[4] 'FAPP'= 'For All Practical Purposes,' first used by J. S. Bell (1990).

[5] Baumann et al (2016) make a similar point, but they assume UO and treat 'collapse' only as a subjective updating. Thus, in their approach, treating improper mixtures as proper mixtures continues, and there is no in-principle resolution of the inconsistency. It can only be assumed to remain 'unmanifest' based on the practical inability of nested observers to communicate, which would not seem to be guaranteed.

But as we saw above, and as Hughes emphasized, the ignorance interpretation of improper mixed states 'simply will not do': (B) *does* logically contradict (A), and the difference is not explainable in terms of ignorance. As we just reminded ourselves, ignorance of the actual outcomes of subsystem measurements does not lead to the same composite-system state as that of the unmeasured subsystems. The marginal on S is an *improper* mixed state that does not license an ignorance interpretation. The inconsistency between the states (A) and (B) is the same as that between (4) and (2).

This inconsistency is so problematic that we can more immediately obtain an interpretive absurdity if we help ourselves, despite the unitary-only assumption, to the idea that Alice and Bob can correctly describe their measurement electrons by outcome-related eigenstates. We don't need to invoke the sophisticated states of the FR scenario in order to get into trouble, as follows. Note that once Alice and Bob have measured their electrons and conferred to assess which of the states in (3) have been actualized, according to them, the electrons are uncorrelated; there is no more entanglement. This means that (according to Alice and Bob) the electrons are no longer in the state (1), so there is no well-defined state of total angular momentum, including no well-defined state of angular momentum in the z-direction. Not only does the state assignment of Alice and Bob disagree with (1) about the probability of a value for the total spin and z-component of spin, it disagrees on whether these physical quantities are even *capable of being defined* for the very same degrees of freedom. This is the sort of inconsistency—more than that, breakdown of theoretical coherence--resulting from overlooking the difference between an improper mixture and a proper mixture.[6]

So, to recap, under UO (which we will question below), *no observer can ever correctly assign to her measured subsystem the eigenstate corresponding to the observed eigenvalue*. Thus, given UO, it is not the case that both observers 'use quantum theory correctly' in the FR scenario. However, since in practice physicists routinely assign outcome-related eigenstates to their measured systems and never find these sorts of inconsistencies, of course this situation is perplexing if one insists on UO. One might settle for this sort of only-FAPP consistency based on the supposed practical impossibility of 'super-observers' like W. But as Bub (2017) correctly notes, "As far as we know, there are no super-observers, but the actuality of a measurement outcome can't depend on whether or not a super-observer turns up at some point."

---

[6] Sudbery (2016) makes a similar point about this basic inconsistency underlying the FR paradox, attributing it to the different observers applying quantum theory in different ways, where one such way includes the PP. But as we have seen, that is a 'band-aid' form of UO; i.e., what we are calling UOPP. The unitary-only assumption is clearly retained by FR in order to create their 'absurd' superposition (B), which is similar to the way in which unfettered linearity leads to the Schrodinger's Cat paradox. It is the unitary-only assumption that leads to the alleged existence of 'absurd' macroscopic superpositions that are never experimentally corroborated but which are crucial for constructing this class of paradoxes. Sudbery (2019) notes that explicit non-unitary collapse theories escape the FR paradox. However, like Nurgalieva and del Rio (2018), he exemplifies such theories by the GRW model, which is not equivalent to quantum theory.

The fundamental problem, of course, is that *unitary-only quantum theory provides no way to get single outcomes out of any measurement* in the first place! This means that the condition S – which is what is empirically observed -- is incompatible with the UO assumption 'out of the starting gate.' The inconsistency inherent in interpreting improper mixtures as proper (epistemic) mixtures is a consequence of that fact. As Tausk (2018) and Bub (2017) have noted, one can maneuver around the FR inconsistency under UO by restricting the assignments of outcome-based eigenstates ('collapsed' states) based on context. The only consistent assignment is to restrict the 'collapsed' degrees of freedom to a particular designated unique and final 'macroscopic' level based on what kinds of observables are being measured. But this gambit comes at a high cost. The negative interpretive consequences are:

(i) the assignment of post-measurement outcome eigenstates to measured systems depends crucially on whether or not there will be a 'super-observer' in the future.
(ii) such assignments are not ontologically licensed by the UO account, so they cannot be ontologically referring; they are fictions as regards the actual physical state of the systems so labeled.[7]
(iii) for cases in which there is a super-observer, one must deny that ordinary experimenters like F are observing the macroscopic level, or one must deny that there is ever a measurement result 'for them,' even as they carry out an ordinary laboratory measurement.

Consequence (i) runs afoul of Bub's correct observation that "the actuality of a measurement outcome can't depend on whether or not a super-observer turns up at some point." This sort of state assignment does not indicate the system's possession of any particular actualized outcome, which leads us to (ii); i.e., under UO any such 'collapsed' state assignment is not ontological – i.e., not merited by the ontological nature of the system to which it is applied. Thus, this contextual assignment of collapsed states amounts to instrumentalism about quantum theory, at least at the designated 'macroscopic' level at which such states are assigned. Moreover, under UO, the fictional 'collapsed' state can be allowed only based on certainty about the non-existence of a future measurement of a particular sort of observable--which, of course, is never attainable. Wallace (2016) has referred to this sort of state assignment as a 'probabilistic' interpretation of the quantum state (in contrast with a 'representational,' i.e., realist, interpretation).[8]

---

[7] The hidden-variable model studied by Sudbery (2016) is subject to (i) but not to (ii) and (iii), since it has a two-level description: (1) the 'pilot' state corresponding to the standard quantum state under UO evolution, and (2) the 'beable' or underlying property state that is assumed to be revealed by the measurement. In that model, (2) has an ontological significance, but it can only be a partial and inadequate description under FR, since it must be interpreted as lacking crucial information concerning possible future measurements for the putative macroscopic superpositions of this class of paradoxes.

[8] Another recent euphemism for instrumentalism about quantum theory is a 'normative' interpretation, since that approach also denies that quantum states refer to any specific physical systems, but rather are features of an instruction manual (quantum theory) for what an observer should expect to experience. This author is

A related approach, discussed by A. Drezet (2018), is to assume that after some time T an observer's memory of a measurement result is 'quantum-erased,' and he remembers only that a definite result occurs, but not the result itself. This supposition is in conflict with the fact that we remember measurement results (or at least, if we are unaccountably forgetful for some reason, there are records of these in the lab which is included in the superposition supposedly presented to W). We never experience being in a state in which we know with certainty that a definite result occurred but have completely forgotten what it is--along with the lab also being 'quantum-erased,' having only a record that a result occurred without any record of the specific result (when it formerly had one). One might argue that this bizarre situation of everyone *and* their labs losing track of a specific result while knowing for sure that a result occurred has never been experienced because the time T is very long. But then this time T depends on whether there will be a particular sort of 'super-observer' in the future. So we are back in the same situation of the future-dependence of quantum state assignments as discussed previously. In this case, even our own memories and the physical conditions of our labs would depend explicitly on the future existence of an appropriate super-observer. [9]

One might argue, following a suggestion of Baumann *et al* (2019), that the FR contradiction would remain 'unmanifest' if super-observers W and ordinary observers F can never communicate. But there seems to be nothing preventing this. For example, W and F could each be modeled as complex molecules with several excitable degrees of freedom. For simplicity, let these degrees of freedom be subject to 2D state spaces. Assume that F has 3 such degrees of freedom labeled by *A,B,C*. Under UO, all a 'measurement' can be is the establishment of a correlation between different degrees of freedom. So the 'measurement' of the initial spin-1/2 system at the level of F can be represented by a correlation between two of F's degrees of freedom *A* and *B,* where *A* plays the role of the spin-½ system prepared in an equal superposition of outcomes 'up' and 'down', and *B* plays the role of detector/memory. *C, a communication degree of freedom,* is unaffected at this stage. According to F, after the 'measurement' of *A* by *B*, he is either in the state

$$|\uparrow\rangle_A \otimes |\uparrow\rangle_B \otimes |0\rangle_C$$

or                                                                                                                                  (5a,b)

---

well aware that some who use these terms don't want to identify their approaches as instrumentalist or antirealist, but according to the *Stanford Encyclopedia of Philosophy,* they are: instrumentalism is "the view that theories are merely instruments for predicting observable phenomena or systematizing observation reports." (Chakravarty, A., 2017). The present author agrees with Wallace's observation that current orthodoxy is "an inchoate attitude to the quantum state, where its dynamics are [assumed to be] always unitary but where it is interpreted either as physically representational or as probabilistic, according to context." (Wallace 2016; my addition in brackets.) Such an approach is fundamentally equivocal about the essential nature of quantum theory, and therefore does not constitute a stable or coherent interpretation of the theory.

[9] Another problem with the account is that the 'quantum-erased' state is represented by the ready state, which cannot be correct: in a ready state, there has been no measurement yet, and therefore no definite result, even if unknown, has been found. So the ready state cannot be the same as the 'quantum-erased' state.

$$|\downarrow\rangle_A \otimes |\downarrow\rangle_B \otimes |0\rangle_C$$

with equal probability. On the other hand, according to W, F ends up in a Bell state with the communication degree of freedom *C* along for the ride:

$$|\Psi\rangle_F = |\Phi^+\rangle|0\rangle_C = \frac{1}{\sqrt{2}}(|\uparrow\rangle_A \otimes |\uparrow\rangle_B + |\downarrow\rangle_A \otimes |\downarrow\rangle_B) \otimes |0\rangle_C \qquad (6)$$

Now let W subject F's degrees of freedom *A* and *B* to a measurement of the 'Bell observable' for which the state $|\Phi^+\rangle$ is an eigenstate. W's experiment could be accompanied by a signal to F as follows. An outcome finding F in the state $|\Phi^+\rangle$ results in a photon being emitted to F to excite his other degree of freedom *C*. If F were to receive that photon for every run of the experiment, the inconsistency would be manifest, since according to F, his probability of receiving the photon, given the mixed state of his A and B degrees of freedom, is 1/2. Now, of course the practical logistics are nontrivial in carrying out this 'super-experiment,' but nothing prevents it, in principle, under UO. Indeed, there is nothing in the UO theory to explain why a molecule could not be considered as 'measuring' its own degrees of freedom through such correlations, since under UO, correlation between degrees of freedom is the necessary and sufficient condition for 'measurement.'[10] Thus, one cannot depend on the idea that the FR contradiction will always remain unmanifest in order to evade the lack of experimentally observed inconsistencies that in principle should result from the use of outcome eigenstates to predict future probabilities for measured systems.

We might also note that assigning fictional 'collapsed' states to what is (under UO) always a subsystem of an entangled state can only be an approximate solution, since interference terms for composite system observables are not absent, but simply small. The ability to detect such tiny interference effects, although unlikely, is at least as likely as the existence of a super-observer like W. So the approximate, FAPP, and equivocal nature of this tactic for achieving consistency with empirical observations cannot be overlooked. If we knew for a fact that Nature obeys unitary-only quantum theory, perhaps we would have to live with this situation. However, UO is not an established fact. If there really is non-unitary collapse in Nature, consistency is restored in an exact manner, without appeal to state assignments that are contingent on the absence of future 'super-observers.' Relinquishing the traditional UO restriction and allowing for real non-unitarity accompanying measurement eliminates this entire class of paradoxes and is straightforwardly consistent with what is already well-corroborated experimental practice:

---

[10] Although measurement is often discussed by reference to 'agents,' this term is an undefined primitive. One cannot object that a molecule or degree of freedom could not be considered an agent, since there is no demarcation between conscious and non-conscious degrees of freedom. Appeals to complexity are question-begging and subject to the hard problem of consciousness, in that complexity alone does not suffice to confer consciousness or intent on an object.

attributing to measured quantum systems the eigenstates corresponding to their outcomes.[11]

3. Reconsidering Real Collapse

In view of the makeshift and *ad hoc* nature of the best-known 'explicit collapse' model, most researchers assume that quantum theory 'really' has only unitary dynamics. This view is underscored by Nurgalieva and del Rio's assertion that 'explicit collapse' theories are 'falsifiable,' where they refer only to the model of Ghirardi-Rimini-Weber (1986). That is, GRW is widely considered as appropriately exemplifying all physical collapse theories, with the attendant assumption that all such theories are crucially different from quantum theory, achieving collapse only through an *ad hoc* nonlinear addition to the Schrodinger equation that forces a collapse that would otherwise never happen. However, there is at least one formulation of quantum theory with a natural, physically well-defined non-unitary process constituting the measurement transition, whose empirical predictions are equivalent to the unitary-only theory (except that it also predicts the observed Boolean event structure that the UO theory fails to predict). That formulation is the Transactional Interpretation. There are other objective collapse theories that do not involve *ad hoc* changes to the Schrödinger equation, such as that of Penrose (1996) and Jabs (2016),

The Transactional Interpretation (TI) is based on the direct-action or 'absorber' theory of fields, which involves an underlying time symmetry of field propagation. Specifically, the direct-action theory uses the time-symmetric propagator instead of the causal Feynman propagator, but the latter is recovered through absorber response, which gives rise to non-unitarity. Thus, in the direct-action theory, response of absorbers is an additional part of the dynamics that can provide a natural physical non-unitarity not found in standard approaches to quantum theory, which neglect the role of absorption in processes constituting 'measurement' interactions. Originated by Cramer (1986), it has now been extended to the relativistic domain by the present author (Kastner 2012, 2016a,b, 2017, 2018, 2019a,b). See Kastner and Cramer (2018) for explicit quantification of the circumstances instantiating non-unitarity, and a physical (as opposed to information-theoretic) derivation of the Born Rule for radiative processes in relativistic TI or 'RTI'.[12]

---

[11] Technically, this applies to non-destructive measurements in which a system is retained for further study in the desired outcome state. But even if the measured system is destroyed, it is common practice to correlate its detected state with another degree of freedom whose prepared state is determined by the previous system's outcome state. An example is an atom placed into a known excited state only if it absorbs a photon in a particular state. Such routine procedures involve assigning outcome-related eigenstates to individual degrees of freedom, which contravenes the unitary-only description in which entanglement is ongoing.

[12] There has sometimes been a stigma attached to the direct-action theory of fields based on its early abandonment by two of its major founders, Wheeler and Feynman (1945, 1949). However, John Wheeler was recently re-advocating the direct-action theory (Wesley and Wheeler, 2003). The theory has some distinct advantages, as discussed in Kastner (2015). The relativistic transactional picture, RTI, decisively overcomes earlier objections such as the Maudlin objection (Maudlin, 2002). For details, see Kastner, 2019a).

It is not the purpose of this paper to present any further details of RTI, but merely to point out that there exist perfectly viable objective collapse interpretations that are empirically equivalent to standard quantum theory (i.e., conform to the Born Rule) and that do not involve *ad hoc* changes to force collapse that would otherwise not occur. Thus, GRW is not an appropriate exemplar of the non-unitary approaches available, and the common practice of citing only GRW, as if that model were indicative of the (high) price to be paid for a non-unitary theory, undermines due consideration of non-unitary approaches as solutions to problems such as that posed by the FR paradox.

Besides the mistaken perception that all collapse models must involve *ad hoc* changes to quantum theory, concerns about real physical quantum state reduction are often based on non-obligatory metaphysical assumptions. One is the conventional assumption that spacetime is the ultimate 'container' for all that exists, and that collapse (therefore) has to be an instantaneous spacetime process, which is in tension with relativity due to its apparent frame-dependence. (cf. Aharonov and Albert 1981). But that is not a mandatory problem: collapse can be understood instead as the process of emergence of invariant spacetime intervals, in a fully covariant, relational (as opposed to substantival) account of spacetime which does not single out any preferred frame (Kastner, Kauffman and Epperson 2018; Kastner 2016a). A related unnecessary assumption is that any account of collapse must predict how and why it would occur 'at a particular time'; i.e., that a deterministic account must be given for the time of collapse. This is arguably a holdover from classical thinking, which must be relinquished if quantum theory is truly an indeterministic theory that correctly describes Nature. Under RTI, the time of collapse is well-defined to within the uncertainty principle, which is an appropriate constraint on the precision of any such prediction (Kastner and Cramer 2018). RTI also yields a natural criterion for the micro/macro transition zone that is consistent with the fact that we never see flagrantly macroscopic superpositions (Kastner 2018).

Another reason to reconsider the existence of collapse is that it resolves the black hole 'information paradox,' which is another sort of paradox that arises only if one insists on unitarity as an inviolable principle. Penrose (1986) has pointed out that if measurement is a non-unitary process, the loss of information into a black hole is not problematic; it is simply a form of measurement.[13] Thus, a genuinely non-unitary form of quantum theory dissolves an entire class of paradoxes based on 'absurd' macroscopic superposition, as well as solving other longstanding problems in physics based on the possibly unnecessary assumption that quantum unitarity should be considered an inviolable physical principle. An additional reason to question UO is that quantum unitarity expresses the conservation of probability, not of an actualized physical quantity. If quantum probability characterizes objective uncertainty about what is to be actualized as a result of measurement, then the relevant sample space consists of un-actualized possibilities. And since not all competing possibilities can be actualized, symmetry must be broken at the spacetime level of actualized events. The latter is the physical correlate of non-unitary quantum state

---

[13] Banks, Susskind and Peskin (1984) have argued that non-unitarity in this context would violate energy conservation, but that argument rests on an apparently overly restrictive assumption and is critiqued in Nikolic (2015).

reduction. For a detailed discussion of this issue, see Kastner, Kauffman and Epperson (2018).

5. Conclusion

The FR paradox illustrates the absurdity of the unitary-only assumption combined with the common practice of ignoring the distinction between a proper mixture (which can be interpreted as epistemic ignorance over an actually possessed state or outcome) and an improper mixture (which does not license the conclusion that a particular state or outcome has been realized). Of course, if one insists on adhering to UO, one is forced to ignore this important distinction for ostensible (FAPP) consistency with the Boolean event structure that is routinely observed in the laboratory but which contradicts the unitary-only model. FR shows us that the measurement problem has not gone away, but instead has returned in spades for unitary-only quantum theory, despite the recent trend of evading it through decoherence-based approaches that assign fictional outcome eigenstates to subsystems that, according to the UO model, have to be in improper mixed states. The FR scenario is thus an important cautionary tale: it illustrates that trying to hold onto the UO assumption in the face of the Boolean structure of observed experimental results is arguably a no-go.[14]

Under UO, *there can be no actual single outcomes*, no ontologically real Boolean structure 'out there in the world'; since that would imply non-unitary collapse with respect to any observable for which more than one outcome is possible given the prepared state, which contradicts the UO assumption. *Under UO, there can be no actual transition from non-Boolean to Boolean event structure*. Thus, a utilitarian, Bohrian account of 'collapse' is necessarily observer-dependent and instrumentalist: collapse is no more than a useful fiction. In such an approach, the attribution of an eigenstate corresponding to a measurement outcome can never be anything but a non-referring instrument that an observer applies for his own convenience. In fact, it's not optional: he *has* to, since his Boolean experience refutes the non-Boolean character of the UO formalism. Yet, since the Bohrian account allows only one 'macrolevel' for any given experimental situation, under certain conditions (the future existence of a super-observer W as in the FR scenario) it must deny that an ordinary human experimenter (analogous to F) observes a definite macro-level outcome—which has never been experimentally corroborated. Meanwhile, in contrast, it is the collapsed, determinate, Boolean description that is always corroborated.

It is of course possible, under UO, to evade the outright FR contradiction and still retain some vestige of the idea that quantum states can be ontologically referring by recourse to a suitable hidden-variable theory. An example is the Bell-Bohm theory discussed by Sudbery (2016), which treats the standard quantum state as a governing

---

[14] Everettians might say that they can keep UO by denying single outcomes, but the noted sentence remains accurate, since what is empirically corroborated *is* a Boolean structure of single outcomes: nobody ever reports seeing all possible outcomes or even more than one possible outcome. Everettians must posit instead a (non-corroborated) set of all possible outcomes, viewed by myriad postulated observer counterparts, in order to retain UO.

'pilot' state, while the 'collapsed' state corresponds to a 'beable' state corresponding to the real property possessed by the system. But in such an approach, F is still not allowed to use the outcome (beable) state to predict future outcomes contingent on the existence of a super-observer. Sudbery notes that in such a theory, pilot state components corresponding to outcomes not realized by an experimenter at the level of F would still be in play to influence future measurements. Thus again, under UO, the observed outcome eigenstate cannot be viewed as a full and accurate description of the detected system.

The appropriate conclusion is that it is unitary-only quantum theory that is suspect, and renewed consideration should be given to the possibility that Nature has real non-unitary quantum state reduction or 'collapse.' For in the latter case, the outcome eigenstates that we always assign in the laboratory and that are always corroborated cease to be fictional; instead they are accurate representations of the physics. When F measures his system, there really *is* an actual outcome. So W will never assign a pure entangled state to F, his apparatus and his lab, and consistency is restored. All the problems pointed to by this class of paradoxes vanish (as well as others, such as the black hole information paradox).

Acknowledgments

I thank Arthur Jabs, James Malley, Tony Sudbery, David Tausk, and two anonymous reviewers for valuable comments.

References

# References

References

Aharonov, Y. and Albert, D. (1981). "Can we make sense out of the measurement process in relativistic quantum mechanics?", *Phys. Rev. D 24*, 359-70.

Araújo, M. (2018). "The flaw in Frauchiger and Renner's Argument," http://mateusaraujo.info/2018/10/24/the-flaw-in-frauchiger-and-renners-argument/

T. Banks, L. Susskind, M.E. Peskin, Nucl. Phys. B 244, 125 (1984).

Baumann, V. , Hansen A., Wolf, S. (2016). "The measurement problem is the measurement problem is the measurement problem." https://arxiv.org/pdf/1611.01111.pdf

Bell, J. (1990). "Against 'Measurement'," *Physics World*. August 1990.

Bub, J. "Why Bohr was (Mostly) Right," https://arxiv.org/pdf/1711.01604.pdf

Chakravartty, Anjan. (2017). "Scientific Realism", *The Stanford Encyclopedia of Philosophy* (Summer 2017 Edition), Edward N. Zalta (ed.), URL = https://plato.stanford.edu/archives/sum2017/entries/scientific-realism/

Cramer J. G. (1986). ``The Transactional Interpretation of Quantum Mechanics.'' *Reviews of Modern Physics 58*, 647-688.

Drezet, A. (2018). "About Wigner Friend's and Hardy's paradox in a Bohmian approach: a comment of 'Quantum theory cannot consistently describe the use of itself,'" preprint, https://arxiv.org/pdf/1810.10917.pdf

Feyerabend, P. K. (1962). "On the Quantum Theory of Measurement." In Korner S. (ed.) (1962). *Observation and Interpretation in the Philosophy of Physics.* New York: Dover, pp. 121-130.

Frauchiger, D. and Renner, R. (2018). "Quantum theory cannot consistently describe the use of itself**,"** *Nature Communications* **9**, Article number: 3711

Ghirardi, G.C., Rimini, A., and Weber, T. (1986). *Phys. Rev. D 34*, 470.

Hughes, R. I. G. (1989). *The Structure and Interpretation of Quantum Mechanics.* Cambridge: Harvard University Press.

Jabs, A. (2016). "A conjecture concerning determinism, reduction, and measurement in quantum mechanics," Quantum Studies: Mathematics and Foundations, Vol.3 (4) 279-292.

Kastner, R. E. (2012). "The Possibilist Transactional Interpretation and Relativity, *Foundations of Physics 42*, 1094-1113. https://arxiv.org/abs/1204.5227